Far-infrared quantum cascade lasers operating in AlAs phonon Reststrahlen band

K. Ohtani, M. Beck, M. J. Süess, and J. Faist

*Institute for Quantum Electronics, ETH Zurich, Auguste-Piccard-Hof 1, 8093 Zurich, Switzerland*

A. M. Andrews, T. Zederbauer, H. Detz, W. Schrenk, and G. Strasser

*Institute of Solid State Electronics and Center for Micro- and Nanostructures, Vienna University of Technology, Floragasse 7, 1040 Vienna, Austria*

**Abstract**

We report on the operation of a double metal waveguide far-infrared quantum cascade laser emitting at 28 μm, corresponding to the AlAs-like phonon Reststrahlen band. To avoid absorption by AlAs-like optical phonons, the Al-free group-V alloy $GaAs_{0.51}Sb_{0.49}$ is used as a barrier layer in the bound-to-continuum based active region. Lasing occurs at a wavelength of 28.3 μm, which is the longest wavelength among the quantum cascade lasers operating from mid-infrared to far-infrared. The threshold current density at 50 K is 5.5 $kA/cm^2$ and maximum operation temperature is 175 K. We also discuss the feasibility that operation wavelength cover the whole spectral range bridging between mid-infrared and terahertz by choosing suited group III-V materials.

A free space wavelength range around 20 - 60 μm lying in the far-infrared (FIR) spectrum between mid-infrared (MIR: ≈ 2 - 20 μm) to terahertz (THz: ≈ 60 – 300 μm) is of an interest for spectroscopic sensing applications such as astrophysics. In contrast to the two neighboring spectral regions of MIR and THz, technologies still remain underdeveloped because of difficulties in building a compact optoelectronics device [1] in the region where incident light strongly couples with lattice vibrations in a solid, leading to a highly dispersive and absorptive optical response with an energy gap called the "Reststrahlen band" [2]

Quantum cascade lasers (QCLs) are semiconductor lasers utilizing intersubband transitions in quantum wells (QWs) [3]. Widely tunable intersubband transition energies together with an engineered optical gain coefficient, enable high power coherent light in the MIR and THz regions [4, 5, 6]. However, it is a challenging task to make them lase in the FIR region because of absorption by polar optical phonons of the group III-V semiconductors: Figure 1 shows a room temperature, normal incident transmission spectrum of a 4 μm thick $In_{0.53}Ga_{0.47}As/Al_{0.48}In_{0.52}As$ active region which is generally employed for MIR and FIR QCLs. Three dips corresponding to the Reststrahlen bands are visible on the spectra. By comparing them to the computed transmission spectra as represented by the dotted line, the three dips are identified to be InAs-like, GaAs-like, and AlAs-like optical phonon Reststrahlen bands, respectively, since those transverse optical (TO) and longitudinal optical (LO) phonon energies in the computation are in good agreement with the reported phonon energies [7, 8, 9]. When expanding QCL operation wavelength from the MIR to the THz, it first reaches the AlAs phonon reststrahlen band. There, generated light is strongly attenuated and absorbed, resulting in an increase of waveguide loss. Such a situation makes QCLs difficult to lase in the Reststrahlen band and was believed to limit the operation wavelength to 24μm [10, 11, 12].

In this work, to extend an operation wavelength into the AlAs phonon reststrahlen band, Al-free GaAsSb barriers are employed in the active region. GaAsSb, with a Sb composition of 49% is lattice matching to the InP substrate, making a type-II QW with the $In_{0.53}Ga_{0.47}As$ layer. Compared to $Al_{0.48}In_{0.52}As$, thicker barriers can be used since it provides a smaller conduction band offset energy (360 meV) and a smaller effective mass (0.045 $m_0$, where $m_0$ is a free electron mass) [13, 14, 15]. To avoid an overlap with the InP phonon Reststrahlen band, a double metal waveguide is employed where the optical field is confined by two thick metal cladding layers, instead of the InP substrate [11, 12].

Figure 2 shows the conduction band diagram of the designed injection/active/injection layers in the active region of an electric field of 22.5 kV/cm. The injection/active layers are based on the bound-to-continuum scheme [16]. As depicted by the red arrow, the optical transition occurs between the weakly bound excited subband and the widely spread ground subband. The emission wavelength is designed to be 28 μm (= 44 meV). The $In_{0.53}Ga_{0.47}As$ well layer with a thickness of 6.4 nm is uniformly doped with Si to supply electrons with a sheet density of $2.6 \times 10^{11}$ cm$^{-2}$. The sample was grown by a solid-source molecular beam epitaxy (MBE) on Fe-doped InP (100) substrates. A 50 nm thick undoped $In_{0.53}Ga_{0.47}As$ bottom buffer layer was first grown, followed by 60 periods of the bound-to-continuum $In_{0.53}Ga_{0.47}As/GaAs_{0.51}Sb_{0.49}$ active region with a total thickness of 4.2 μm. The growth was completed by a 30 nm thick $n$-$In_{0.53}Ga_{0.47}As$ top contact layer ($n = 2\times10^{18}$ cm$^{-3}$). The barrier thicknesses were adjusted, compared to the nominal value, by a factor of 10-20%, because our Schrödinger and Poisson solver underestimates wavefunction couplings between QWs [17]. After MBE growth, the wafers were processed into Au double metal waveguide ridge laser structures by a standard photolithography and a wet chemical etching. The $n$-In$_{0.53}$Ga$_{0.47}$As top contact layers were removed to decrease the waveguide loss.

Figure 3 (a) shows temperature dependent light output – current – voltage characteristics of the laser device with a ridge width of 30 μm and a cavity length of 1.0 mm. At the temperature of 10 K, lasing was observed with a threshold current density ($J_{th}$) of 5.5 kA/cm$^2$. The maximum current density ($J_{max}$) was ≈ 9.0 kA/cm$^2$. The peak output power of 0.1 mW was recorded by a calibrated Si bolometer. High voltage drop was observed on the transport curves due to the formation of a Schottkey barrier by a lack of the $n$-In$_{0.53}$Ga$_{0.47}$As top contact layer. Figure 3 (b) shows emission spectra at 10 K. The laser center emission energy is 43.7 meV, corresponding to the wavelength of 28.3 μm. When the laser is pumped close to the negative differential resistance region, the laser emission spectrum spans from 26.6 to 28.5 μm, reflecting a broad gain spectra. As depicted in Figs. 3 (a) and (c), the maximum operation temperature ($T_{max}$) was 175 K, which is lower than $T_{max}$ (= 240 K) of our previous λ = 24 μm In$_{0.53}$Ga$_{0.47}$As/Al$_{0.48}$In$_{0.52}$As QCL. The lower $T_{max}$ is attributed to the smaller current density dynamic range, the smaller $J_{max}$, since the characteristic temperature ($T_0$ = 234 K) of $J_{th}$ by fitting from 110 to 170 K is comparable to $T_0$ of the 24 μm In$_{0.53}$Ga$_{0.47}$As/Al$_{0.48}$In$_{0.52}$As QCL [12].

Figures 4 show (a) low temperature $J_{th}$ as a function of laser emission energies and (b) normal incident transmission spectrum. For comparison, data of In$_{0.53}$Ga$_{0.47}$As/GaAs$_{0.51}$Sb$_{0.49}$ QCLs operating at other wavelengths (26.5 μm and 27.2 μn) and In$_{0.53}$Ga$_{0.47}$As/Al$_{0.48}$In$_{0.52}$As QCLs having a similar gain coefficient are added. $J_{th}$ of In$_{0.53}$Ga$_{0.47}$As/GaAs$_{0.51}$Sb$_{0.49}$ QCLs is almost constant (≈ 5.0 kA/cm$^2$) because of no absorption band in this energy range. However $J_{th}$ of the In$_{0.53}$Ga$_{0.47}$As/Al$_{0.48}$In$_{0.52}$As QCLs increases when the laser emission energy approaches the AlAs phonon energy since the lasing energy starts to overlap with the AlAs phonon Reststrahlen band, resulting in an increase of the waveguide loss. The longest operation wavelength (26.3 μm) of In$_{0.53}$Ga$_{0.47}$As/Al$_{0.48}$In$_{0.52}$As QCLs is in close agreement with our previous prediction (27.3 μm), based on the estimated gain coefficient (11.5 cm$^{-1}$/kA) and the computed waveguide loss by phonon absorption [12]. Here we apply the same argument to discuss a limitation to the longest operation wavelength of In$_{0.53}$Ga$_{0.47}$As/GaAs$_{0.51}$Sb$_{0.49}$ QCLs. Figure 4 (c) shows computed total loss of the Au double metal waveguide with an In$_{0.53}$Ga$_{0.47}$As/GaAs$_{0.51}$Sb$_{0.49}$ active region thickness of 4.0 μm. The dielectric constant based on the effective medium approximation is employed [18] since it reproduces the experimental transmission curve as depicted by the dashed line in Fig. 4 (b). Computed optical mirror losses (10-15 cm$^{-1}$) are added although they are smaller than the waveguide loss (> 50 cm$^{-1}$). The gain coefficient is assumed to be independent of wavelength, and comparable to that (11.5 cm$^{-1}$/kA) of In$_{0.53}$Ga$_{0.47}$As/Al$_{0.48}$In$_{0.52}$As QCLs [12], because the gain coefficients computed by the density matrix method [19] and measured $J_{th}$ are similar. Taking the same $J_{max}$ (9 kA/cm$^2$) as the present 28 μm QCL gives the estimated longest operation wavelength of 32 μm (39 meV). The value is shorter than our expectation since there is an energy spacing (≈ 7 meV) to the GaAs-like TO phonon (32 meV). This is due to the large absorption tail of the GaAs optical phonon.

To further extend the operation wavelength of QCLs it reaches the Reststrahlen band of GaAs and InAs, which are used as well layers in all the state-of-art QCLs [3, 4, 5, 6, 13, 14, 20, 21, 22, 23]. To avoid excitations of those phonons, an alternative material for well layers would be InSb because the phonon energies (22.3 meV for TO phonon, and 23.8 meV for LO phonon [24, 25]) are smaller than the InAs and the GaAs phonons. InSb would be attractive for QCL materials since it has the smallest effective mass among the group III-V semiconductors, expecting a large optical gain coefficient [26]. As depicted in fig. 4 (d), the comparable gain coefficient (11.5 cm$^{-1}$/kA) with the similar $J_{max}$ (9 kA/cm$^2$) predicts the longest operation wavelength around 41 μm (= 30 meV). To reach the high frequency THz region where THz GaAs QCLs start to work [27], InP QWs with lattice matching AlAs$_{0.56}$Sb$_{0.44}$ barriers would be a good candidate: thanks to the larger optical phonon energies of InP (TO: 38 meV, LO: 43 meV) [24], and AlAsSb (TO: ≈ 42 meV, LO: ≈ 46 meV) [28], it is more transparent compared to GaAs/AlGaAs QWs as shown in fig. 4 (d). At the boundary region between InP and InSb QWs, as shown by the black arrow in fig. 4 (d), waveguide loss of the double metal waveguide can be decreased by 15% if employing a thicker active region (≈ 10 μm) which is comparable to that of THz QCL. Hence it could be possible that the operation wavelength covers a whole spectral range bridging between mid-infrared and terahertz by choosing the suited combination of the group III-V materials.

In conclusion, we have demonstrated far-infrared quantum cascade lasers emitting at the 26 – 28 μm wavelength range, corresponding to the AlAs phonon Reststrahlen band, by replacing AlInAs barriers with GaAsSb ones. $J_{th}$ at 10K was 5.5 kA/cm$^2$ and $T_{max}$ was 175 K. The operation wavelength was 28.4 μm, which is so far the longest wavelength in far-infrared and mid-infrared QCLs. We also discussed group III-V QCL material systems to cover the whole spectral region between MIR to THz and found that InSb and InP-based QWs are requisite materials.

This work was supported by ETH Zurich with the FIRST cleanroom facility, the collaborative research center 956 (SFB956) funded by Deutsche Forschungsgemeinschaft (DFG), the Austrian Science Funds (FWF) within SFB project F49-09 NextLite, the frame work of Doctoral School "Building Solids for Function" (project W1243) as well as through the Gesellschaft für Mikro- und Nanoelektronik (GMe).

**Figure captions**

Figure 1   Room temperature, normal incident transmission spectrum of a 4 μm thick $In_{0.53}Ga_{0.47}As/Al_{0.48}In_{0.52}As$ far-infrared quantum cascade laser active region. The active region film was glued on the Si substrate with the epoxy. The measurement was done with a Fourier transform infrared spectrometer equipped with a Mylar beam splitter and a room temperature Deuterated Tri Glycine Sulfate (DTGS) detector. The dashed line represents simulated transmission spectra. The phonon damping constant is assumed to be 8 $cm^{-1}$, being independent on the materials.

Figure 2   Conduction band diagram of injection/active/injection layers in the $In_{0.53}Ga_{0.47}As/GaAs_{0.51}Sb_{0.49}$ active region at an electric field of 22 kV/cm. Effective masses of $In_{0.53}Ga_{0.47}As$ (0.043 $m_0$) and $GaAs_{0.51}Sb_{0.49}$ (0.045 $m_0$) are used in the computation. The conduction band offset energy is 0.36 eV. The injection and active layers are repeated 60 times, corresponding to the total thickness of 4.2 μm. One of the well layers in the injection layer is doped Si to be $2.6 \times 10^{11}$ $cm^{-2}$.

Figure 3   (a) Light output – current – voltage characteristics as a function of temperature. The laser device has a ridge width of 30 μm and a cavity length of 1.0 mm. Current pulses with a pulse width of 100 ns and a repetition rate of 450 Hz were used. (b) Laser emission spectra biasing below (1.85 A) and above (2.75 A, the inset) negative differential resistance region. (c) Threshold current densities as a function of temperature. $T_0 = 213$ K was obtained by fitting an empirical equation of threshold current density. The maximum operation temperature was 175 K.

Figure 4   (a) Room temperature normal incident transmission spectra and (b) Threshold current densities as a function of laser emission energy of $In_{0.53}Ga_{0.47}As/GaAs_{0.51}Sb_{0.49}$ and $In_{0.53}Ga_{0.47}As/Al_{0.48}In_{0.52}As$ QCLs equipped with the Au double metal waveguide. (c) Computed total loss of the Au double metal waveguide based on $In_{0.53}Ga_{0.47}As/GaAs_{0.51}Sb_{0.49}$ QCLs. The active region thickness is 4.0 μm. The dashed line represents the estimated maximum gain coefficient. (d) Computed total loss of the Au double metal waveguide of InP/AlAsSb and InSb/AlInSb active regions with the thickness ratio ($L_b^{total}/(L_w^{total} + L_b^{total}) = 15\%$) where $L_w^{total}$ and $L_b^{total}$ shows the total thickness of well and barrier. The active region thickness is 4.0 μm. The phonon damping constant is assumed to be 8 $cm^{-1}$, being independent on the materials.

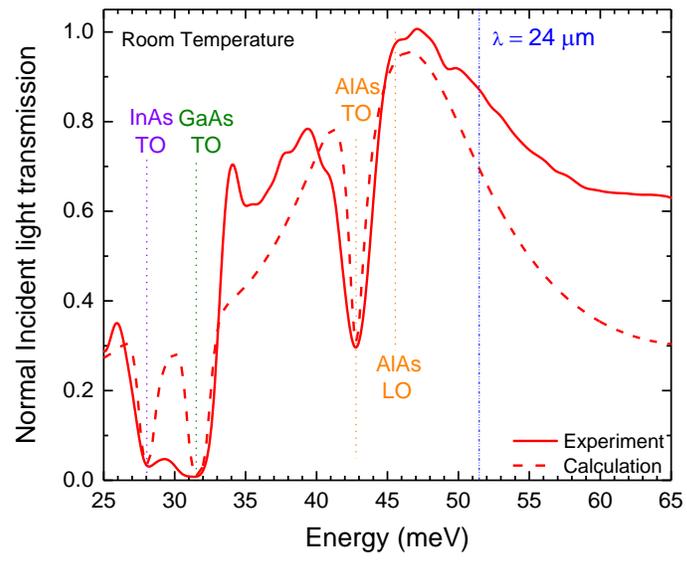

Fig. 1

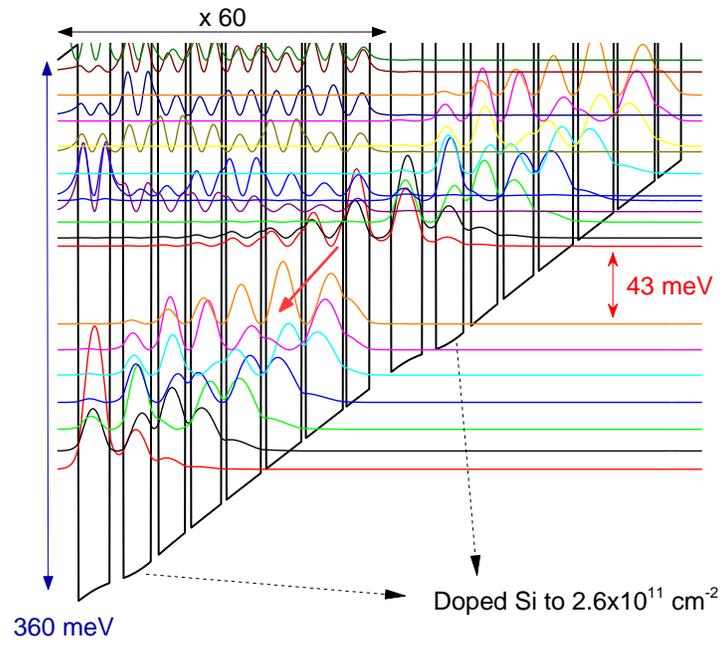

Figure 2

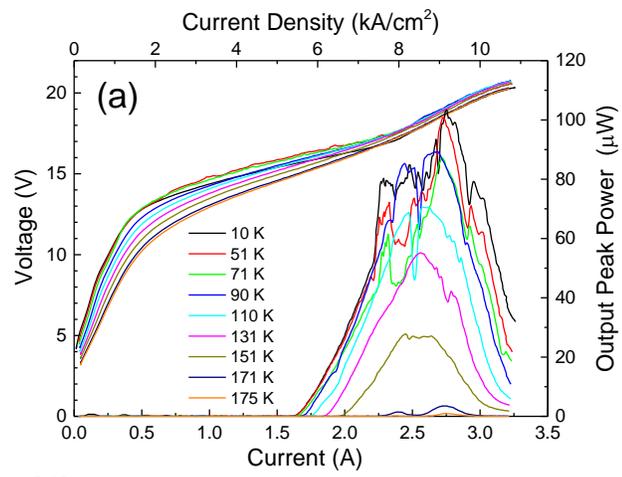

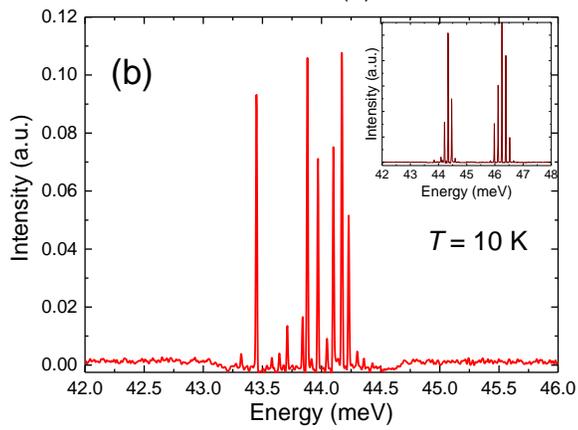

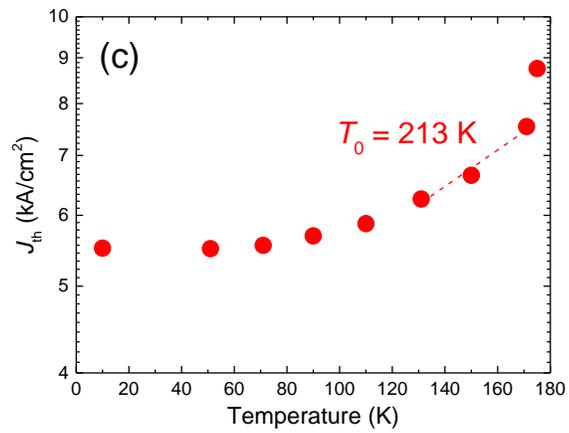

Figure 3

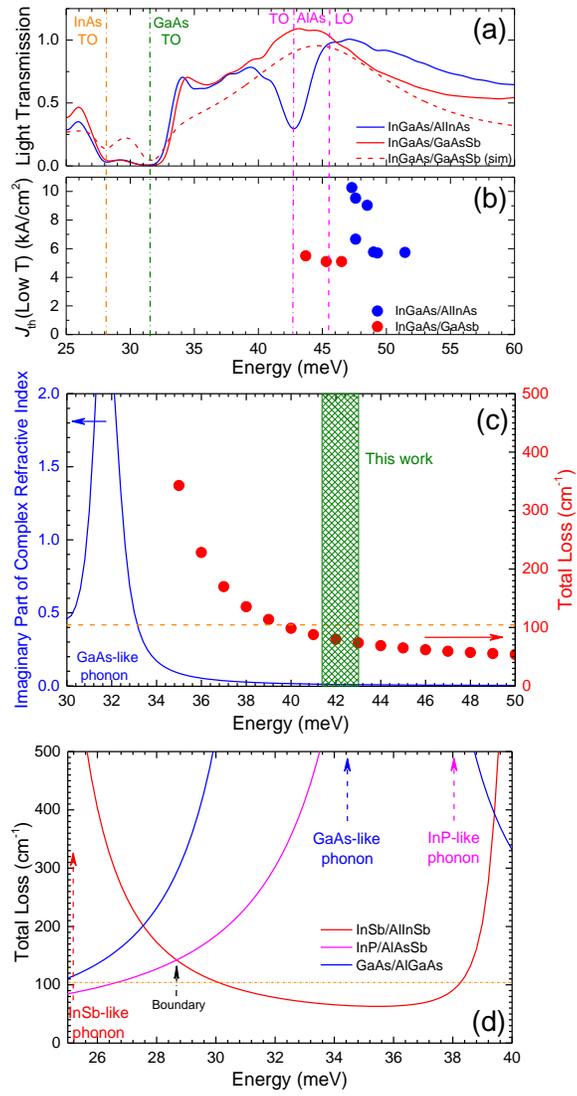

Figure 4

## Supplementary Information

- **Characterization of wavefunction coupling between InGaAs/GaAsSb quantum wells**

A wavefunction coupling between quantum wells (QWs) determines carrier transport, lasing energy, and optical gain of QCLs. First we evaluate it on InGaAs/GaAsSb QWs since it is a crucial parameter to design the active region structure. Coupled QWs (CQWs) having two different intersubband transitions as depicted in Fig. S1 (a) are suited for this purpose because energies and oscillator strengths of those transitions are dependent on how large far wavefunctions penetrate into the adjacent wells. Here three CQWs samples (the left well thickness is 6 nm and the right well thickness is 2.3 nm) having different middle GaAsSb barrier thicknesses ($d$ = 1.4, 1.8, and 2.2 nm thick, respectively) are prepared. Fig. S1 (b) shows intersubband absorption spectrum of one of those three samples. Two absorption peaks are observed and identified that the lower energy absorption peak is attributed to the intersubband transition from E1 to E2 while the higher energy one is associated with the transition between E1 and E3. Fig. S1 (c) depicts integrated absorption intensity ratios ($I_{12}/I_{13}$) of those two peaks. $I_{12}/I_{13}$ is here given by $I_{12}/I_{13} = (E_{12}/E_{13}) \times (z_{12}/z_{13})^2$. The dotted line represents the computed $I_{12}/I_{13}$ by the Schrödinger and Poisson solver taking into account of the depolarization shift. Since the energy ratio of $E_{12}/E_{13}$ from the experiments is in close agreement with the computed ones within 5% (Fig. S1 (d)), the large discrepancy as seen in Fig. S1 (c) is attributed to the term of $(z_{12}/z_{13})^2$. In those CQWs, a thicker middle barrier makes the states ($E_1$, $E_2$, and $E_3$) localized, leading to an increase of $z_{13}$ and a reduction of $z_{12}$. Hence, the larger $(z_{12}/z_{13})^2$ experimentally observed reveals that the actual wavefunctions must penetrate into the adjacent wells. In order to compensate the difference, we increase the barrier thicknesses, in particular, for thicker barriers, compared to the nominal value by a factor of 10-20% as follows:

(Designed structure) **4.0**/5.4/**0.55**/8.6/**0.75**/8.2/**0.75**/8.1/**0.8**/7.1/**0.9**/6.1/**1.4**/6.4/**2.6**/7.2

(Grown structure) **4.8**/5.4/**0.55**/8.6/**0.75**/8.2/**0.75**/8.1/**0.85**/7.1/**1.05**/6.1/**1.6**/6.4/**3.0**/7.2

Here the layer sequence is started from injection barrier and the GaAsSb barrier is in bold.

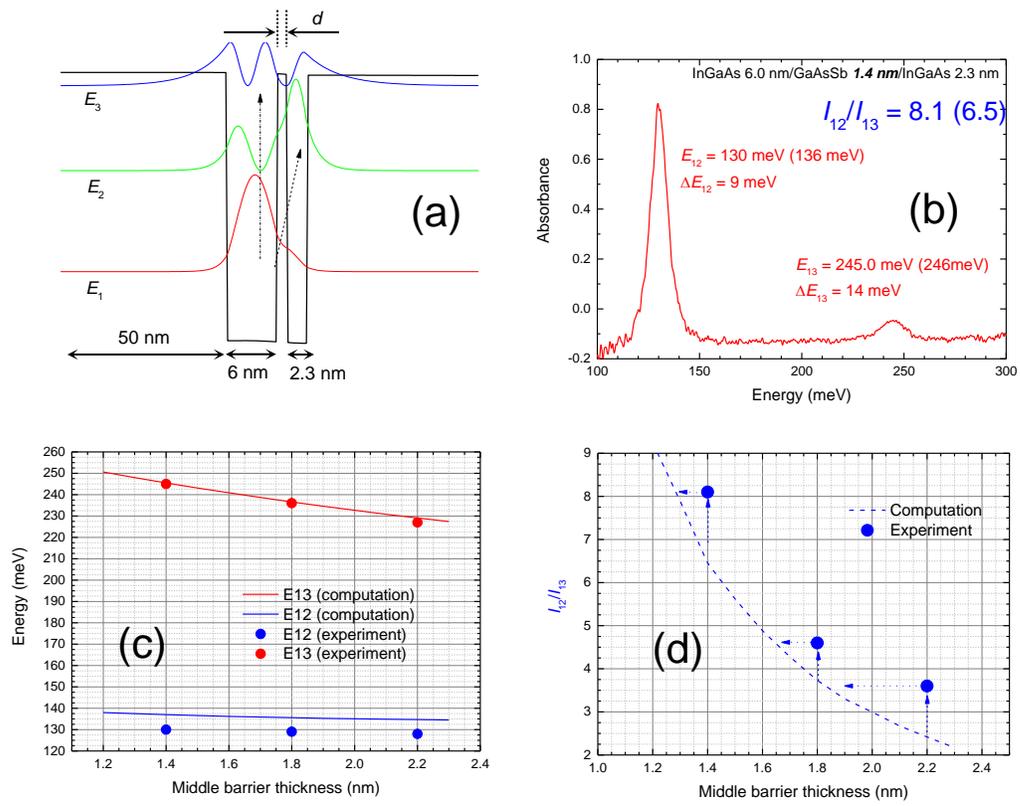

Fig. S1